\documentclass[lettersize,journal]{IEEEtran}
\usepackage{amsmath,amsfonts}
\usepackage{soul}
\usepackage{amssymb}
\usepackage{physics}
\usepackage{algorithm}
\usepackage{algpseudocode}
\usepackage{array}
\usepackage{bm}
\usepackage{color}
\usepackage{booktabs}
\usepackage{subcaption}
\usepackage{mathtools}
\usepackage{xcolor}
\usepackage{tabularx}
\usepackage{tikz}
\usepackage{multirow}
\usetikzlibrary{arrows.meta, positioning, shapes.geometric, fit, backgrounds, calc, chains, 3d}
\usepackage{makecell}
\usepackage{textcomp}
\usepackage{stfloats}
\usepackage{url}

\usepackage{verbatim}
\usepackage{graphicx}
\usepackage{cite}
\usepackage{pdfpages}
\usepackage{tikz-3dplot}
\usepackage{relsize}
\usepackage[hidelinks]{hyperref}
\usepackage{siunitx}
\usepackage{eqparbox}
\usepackage{courier}
\definecolor{mycustomcolor1}{rgb}{0.6627, 0.1412, 0.1255} 
\definecolor{mycustomcolor2}{rgb}{0.8196, 0.5686, 0.2431}
\definecolor{mycustomcolor3}{rgb}{0.0549, 0.1922, 0.3765}
\definecolor{mycustomcolor4}{rgb}{0.0745, 0.3255, 0.5647}
\definecolor{mycustomcolor5}{rgb}{0.7882, 0.3490, 0.3725} 
\definecolor{mycustomcolor6}{rgb}{0.2157, 0.1843, 0.1843}
\definecolor{mycustomcolor7}{rgb}{0.9843, 0.8078, 0.2275}
\definecolor{mycustomcolor8}{rgb}{0.8784, 0.8157, 0.7216}

\newcolumntype{C}[1]{>{\centering\arraybackslash}m{#1}}
\newcolumntype{L}[1]{>{\centering\arraybackslash}m{#1}}
\algrenewcommand\alglinenumber[1]{\small\ttfamily\textcolor{black}{#1}}
\algrenewcommand\algorithmicrequire{\textbf{\small\ttfamily Input:}}
\algrenewcommand\algorithmicensure{\textbf{\small\ttfamily Output:}}
\algrenewcommand\algorithmiccomment[1]{\hfill\#\ \eqparbox{COMMENT}{\small\ttfamily #1}}
\begin{document}
\newcommand{\orcidiconAbk}{\href{https://orcid.org/0009-0006-1187-7782}{\includegraphics[scale=0.1]{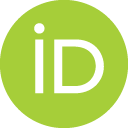}}}
\newcommand{\orcidiconOba}{\href{https://orcid.org/0000-0003-2523-3858}{\includegraphics[scale=0.1]{Figures/orcidID128.png}}}
\title{Acoustic, VOC, and Multimodal Stress Source Localization in the Internet of Plants}
\author{Ahmet B. Kilic\orcidiconAbk,~\IEEEmembership{Student Member,~IEEE}, 
        and Ozgur B. Akan\orcidiconOba,~\IEEEmembership{Fellow,~IEEE}                
        \thanks{The authors are with the Center for neXt-generation Communications (CXC), Department of Electrical and Electronics Engineering, Koç University, Istanbul, Turkey (e-mail: \{ahmetkilic20, akan\}@ku.edu.tr).}
        \thanks{Ozgur B. Akan is also with the Centre for neXt Communications (CXC), Department of Engineering, University of Cambridge, CB3 0FA Cambridge, U.K (e-mail: oba21@cam.ac.uk).}
\
	    \thanks{This work was supported in part by the AXA Research Fund (AXA Chair for Internet of Everything at Ko\c{c} University).}
}
	
\maketitle
\begin{abstract}
The Internet of Plants (IoP) treats distributed plant networks as bio-sensing
infrastructure for environmental monitoring, but spatial localization of stress
sources within such networks remains unaddressed. Plant stress signals have
fundamentally different spatial dynamics: acoustic emissions propagate
omnidirectionally and independently of wind, while volatile organic compound
(VOC) plumes are narrow and advection-dominated. We propose a two-stage,
coarse-to-fine localization pipeline for a network of ``agent plants''---bio-hybrid
sensing nodes embedded in the canopy. Stage~1 localizes the source via
time-difference-of-arrival (TDOA) multilateration on acoustic time-of-arrival (ToA) readings; Stage~2 refines this estimate using a closed-form, steady-state
Green's function model of VOC dispersion. A VOC informativeness gate and an
inverse-variance fusion rule combine the two estimates according to their
across-trial reliability, with graceful degradation to the TDOA-only estimate
when no informative VOC signal is detected. We evaluate TDOA-only, VOC-only,
and fused approaches on a new open-source dataset of 52 scenarios generated via
a finite-volume advection-diffusion solver and a ray-based acoustic attenuation
model. Across network densities of 1 to 50 agent plants, TDOA multilateration
achieves sub-meter mean absolute error (MAE) once three or more agents are within acoustic range,
far outperforming VOC-only localization (MAE $> 3$~\si{\meter} at all densities)
and the single-anchor fallback used below this threshold. Fusion differences from the TDOA-only estimate are small and statistically
indistinguishable from noise in most cases. The pipeline is robust to physical parameter perturbations, ToA
noise, the VOC gate threshold, and the bounding radius. TDOA localization is
deployable with current acoustic hardware, whereas VOC localization remains a
forward-looking capability pending advances in compact biochemical sensors.
\end{abstract}
\begin{IEEEkeywords}
Internet of Plants, Plant Stress Localization, Acoustic Emissions, Volatile Organic Compounds, Multimodal Localization, Agent Plants, Modality Comparison
\end{IEEEkeywords}

\section{Introduction}
\IEEEPARstart{P}{lants} are not passive organisms. Under stress
conditions such as drought, herbivory, or mechanical damage, they emit
a range of physically distinct signals that propagate through the
surrounding environment. These include volatile organic compounds
(VOCs) released from leaf surfaces, airborne ultrasonic pulses
generated by cavitation events in the xylem, and electrical potentials
propagating along stem tissues~\cite{kilic2025}. This coexistence of
multiple signaling modalities has attracted growing interest from the
communications and networking community.

A body of work now frames inter-plant signaling within molecular
communication theory. VOC transport has been characterized as a
diffusion channel, and end-to-end mathematical models of stress
communication have been derived, including channel models for specific
volatile species~\cite{Kilic2026EndToEnd, Merdan2026GLV} and
modulation schemes for odor-based information transfer~\cite{kilic2025}. 
Plant acoustic emissions have been studied as a complementary airborne channel.
Experiments show that stress type and severity can be classified from
ultrasonic pulse signatures recorded at distances of several
meters~\cite{Khait2023}, and the first end-to-end acoustic
communication framework for plants has been
developed~\cite{Merdan2025Acoustic}. Electrical signaling along plant
stems has also been characterized as a guided wired
channel~\cite{Volkov2018}.

Building on this theoretical foundation, the Internet of Plants (IoP)
is a paradigm that treats distributed plant
networks as self-organizing bio-sensing infrastructure capable of
environmental monitoring without dedicated hardware~\cite{kilic2025}.
Within this paradigm, ``agent plants'' have been introduced
as a bio-hybrid sensing architecture~\cite{bilgen2024}. An agent plant
is a designated plant equipped with minimal
electronics to read out the VOC and acoustic signals it naturally
receives from its neighbors. Because agent plants are
indistinguishable from the rest of the canopy, they can be deployed at
scale across a growing area without the prohibitive infrastructure
costs of conventional sensor networks, forming a spatially
distributed, multimodal sensing array embedded within the field itself.

However, sensing that a neighboring plant is stressed is only the
first step; targeted agricultural intervention requires knowing exactly
where that plant is located. The detection and classification of plant
stress signals are well established~\cite{midzi2022stress, Copolovici2011, Khait2023},
but the spatial localization of the stress source within an IoP framework remains
unaddressed. Gas and odor source localization has been studied in robotic
olfaction using Bayesian plume tracking~\cite{GasAdvection2019} and
Physics-Informed Neural Networks for atmospheric
pollution~\cite{ChuprovPINN2025, PhysicsGSL2024}. Acoustic microphone array
methods based on time difference of arrival and beamforming provide robust
localization in general environments~\cite{Grumiaux2022SSL}. None of these
approaches operate in the constrained bio-sensing setting of the IoP, where
sensing nodes are fixed living organisms, multiple modalities can be fused,
and the physical structure of plant signaling channels dictates performance.

This gap is compounded by a physical asymmetry between the available
signaling channels: VOC transport is advection-dominated under realistic
airflow, producing a narrow, directional plume that leaves sensors outside
it with effectively zero signal regardless of proximity, whereas acoustic
emissions propagate omnidirectionally at the speed of sound so any sensor
within the detection radius receives an informative signal. The two channels
therefore carry different spatial information, motivating a precise
characterization of when each contributes and how they can be fused. We
focus on these two airborne modalities because, unlike electrical signaling,
they propagate through the surrounding medium and can therefore reach
sensors at greater distances from the source.

To address this, the present work systematically investigates acoustic-only,
VOC-only, and multimodal fusion localization, extending our preliminary
findings~\cite{kilic_akan_2025} with TDOA multilateration, an entirely new VOC localization stage and
fusion mechanism, a larger open-source dataset, and a systematic
three-modality evaluation.

This work makes four primary contributions:

\begin{enumerate}
    \item A physics-based, open-source dataset of 52 scenarios, generated using
    a custom finite-volume advection-diffusion solver for
    VOC transport and a ray-based attenuation model for acoustics. Because no 
    empirical multimodal dataset combining VOC and acoustic plant stress measurements 
    is publicly available, this dataset provides the critical missing intermediate step 
    between simplified channel models and full field deployment, covering multiple 
    source positions and wind conditions.
    \item A geometry-driven, greedy QR agent plant selection algorithm that places
    sensing nodes based solely on planting area geometry, without requiring prior 
    measured signal data.
    \item A coarse-to-fine localization pipeline that couples a closed-form
    TDOA multilateration estimator with a differentiable, physics-based
    inverse solver based on a steady-state advection-diffusion Green's
    function. A VOC informativeness gate and an inverse-variance fusion rule
    link the two stages, so the system falls back safely to the TDOA estimate
    when the VOC signal is unavailable or uninformative.
    \item A systematic evaluation across network densities ($N=1$ to $50$) showing
    that TDOA multilateration is a robust, sub-meter-accurate performance
    floor once three or more agents are within acoustic range, and that
    multimodal fusion differences from TDOA-only are statistically indistinguishable from noise in most cases. The
    pipeline is also shown to be robust to physical parameter perturbations,  ToA noise, the VOC gate threshold, and the choice of bounding radius.
\end{enumerate}

The remainder of this paper is organized as follows.
Section~\ref{sec:dataset} describes the simulation environment and
dataset. Section~\ref{sec:agents} presents the agent plant selection
methodology. Section~\ref{sec:pipeline} details the coarse-to-fine
localization framework. Section~\ref{sec:eval} reports the evaluation
results, and Section~\ref{sec:conclusion} concludes the paper. The dataset and all evaluation code are available on GitHub.\footnote{\url{https://github.com/Aburakkilic/Acoustic-VOC-and-Multimodal-Stress-Source-Localization-in-the-Internet-of-Plants}}

\section{Simulation Environment}
\label{sec:dataset}
To evaluate localization performance across these different
signaling channels, we developed a custom physics-based simulation environment.
It couples two independent numerical solvers---a finite-volume
advection-diffusion solver for VOC transport and a ray-based attenuation model
for acoustic propagation---operating over a shared spatial domain and plant
geometry to ensure physical consistency.

\subsection{Simulation Domain}
The simulation domain reflects realistic high-density
agricultural conditions while remaining computationally tractable for systematic
evaluation. The environment represents a rectangular planting area measuring 
$15 \times 20 \times 3$~\si{\meter} in the $x$, $y$, and $z$ directions, 
respectively. Plants are arranged on a regular grid with 0.75~\si{\meter} spacing. 
Each plant is modeled as an upright cylinder (radius 0.05~\si{\meter}, 
height 1.0~\si{\meter}), placed at grid positions from 0.75 to 14.25~\si{\meter} 
in $x$ and 0.75 to 19.5~\si{\meter} in $y$. This arrangement yields a 
$19 \times 26$ grid comprising 494 plant positions. 

In every scenario, one plant acts as the active stress source, while the 
remaining 493 serve as candidate receiver positions for agent plant deployment. 
Key simulation parameters are summarized in Table~\ref{tab:simparams}.

\begin{table}[t]
\centering
\caption{Simulation Parameters}
\label{tab:simparams}
\renewcommand{\arraystretch}{1.2}
\begin{tabular}{llr}
\toprule
Parameter & Value & Reference \\
\midrule
Domain $(x \times y \times z)$ & $15 \times 20 \times 3$~m & -\\
Plant spacing                  & 0.75~m                    & -\\
Grid resolution                & 0.05~m                    & -\\
Simulation duration            & 120~s at 1~s intervals    & -\\
$D_\text{VOC}$                 & $1.6\times10^{-5}$~m$^2$/s & \cite{burgess2024diffusion} \\
VOC surface flux               & 0.5~nmol/m$^2$/s          & \cite{Copolovici2011} \\
SPL reference level            & 65~dBSPL at 0.1~m         & \cite{Khait2023} \\
Acoustic attenuation           & 1.6~dB/m                  & \cite{Khait2023} \\
\bottomrule
\end{tabular}
\end{table}

\subsection{VOC Transport Model}
Within this 3D canopy, VOC dispersion is governed by the 
advection-diffusion equation:
\begin{equation}
\frac{\partial C}{\partial t} + \mathbf{u} \cdot \nabla C
= D_\text{VOC}  \nabla^2 C + Q \, \delta(\mathbf{x} - \mathbf{x}_s),
\label{eq:advdiff}
\end{equation}
where $C$ is the molar concentration, $\mathbf{u} = (u_x, u_y, 0)$
is the horizontal wind vector, $D_\text{VOC}$  is the molecular diffusivity,
$Q$ is the emission rate, and $\mathbf{x}_s$ is the source location.
Wind advection acts exclusively in the horizontal plane, leaving vertical 
transport purely diffusive. The stress event is modeled by distributing the 
source emission onto the computational grid at the plant apex 
($z = 1.0$~\si{\meter}) via trilinear interpolation.

The wind field $\mathbf{u}$ is constant, horizontal, and spatially uniform
across the domain, with no turbulent fluctuations and no vertical component,
representing a laminar or low-turbulence limiting case adopted for
computational tractability and reproducibility.

To solve this system, we implemented a custom finite-volume explicit Euler solver
using vectorized NumPy operations. The solver applies a first-order
upwind scheme for advection and a standard second-order centered scheme for
diffusion. Open (outflow) boundary conditions are applied at the four lateral
walls and the ceiling, allowing VOC mass to exit the domain via both advection
and diffusion (assuming $C=0$ outside the domain), while no-flux conditions are
enforced at the ground floor ($z=0$) and at any face shared with a plant
cylinder surface, by zeroing the flux across that face. Numerical stability is
maintained by evaluating the Courant--Friedrichs--Lewy (CFL) condition at every
timestep.

To capture the vertical concentration profile within the canopy layer, 
the VOC concentration is sampled at eight azimuthal points around 
each receiver's perimeter at three distinct heights (0.75, 1.0, and 
1.25~\si{\meter}). Averaging these eight perimeter samples yields a single 
concentration value per height, per timestep, producing a final dataset 
field of shape $(493, 3, 120)$ (receivers $\times$ heights $\times$ timesteps).

\subsection{Acoustic Propagation Model}
In parallel with the VOC dispersion, a separate acoustic model computes the
sound pressure level (SPL) received at each plant position. Following the 
empirical emission characteristics reported in~\cite{Khait2023}, the 
received SPL at distance $r$ from the source is:
\begin{equation}
\text{SPL}(r) = 65 - 20 \log_{10}\!\left(\frac{r}{0.1}\right)
- 1.6\,(r - 0.1),
\label{eq:spl}
\end{equation}
where the first logarithmic term captures geometric spreading, and the second 
linear term accounts for atmospheric attenuation. Both the source and receivers 
are evaluated at a height of $z = 1.0$~\si{\meter}.

The model traces both a direct ray and a ground-reflected ray (via an image
source at $z = -1.0$~\si{\meter}) to each receiver. To simulate incoherent
reflections within a densely planted canopy, the reflection phase is drawn
independently and uniformly from $[0, 2\pi]$ for each perimeter point.
Physical obstructions are accounted for via a vectorized 3D ray-cylinder
intersection test. If all direct and reflected rays to a receiver are obstructed
by intervening plants, a sentinel value of $-999$ is assigned, indicating
a complete loss of line-of-sight. Otherwise, the pressures at the eight
perimeter points are averaged and converted to dBSPL, yielding an acoustic
dataset field of shape $(493, 1)$.

In addition, for each receiver the minimum propagation time across the
unobstructed direct and reflected rays is recorded as the time of arrival
(ToA), computed using a sound speed of $c = 343$~\si{\meter\per\second},
yielding a corresponding ToA field of shape $(493, 1)$.

\subsection{Scenario Design}
To cover the full aerodynamic and geometric variance a
deployed network would face, we generated 52 distinct scenarios.

Four source locations (snapped to the nearest grid position) were selected to
represent different topological challenges: a central position at
$(7.5, 9.75)$~\si{\meter}, two lateral positions at $(3.0, 9.75)$
and $(12.0, 9.75)$~\si{\meter}, and a near-boundary corner position
at $(1.5, 1.5)$~\si{\meter}. The corner source represents a worst-case geometric
constraint, as its proximity to the boundary severely limits the number of agent
plants that can surround it, making localization coverage much harder to achieve.

For each location, thirteen wind conditions were simulated: four horizontal
directions ($+x$, $+y$, $+45^\circ$, and $+135^\circ$) at three speeds (0.2, 0.6,
and 1.0~\si{\meter\per\second}, representing low, moderate, and high ventilation),
plus a pure-diffusion baseline with zero wind. This matrix yields
$4 \times (4 \times 3 + 1) = 4 \times 13 = 52$ total scenarios.

\section{Agent Plant Selection}
\label{sec:agents}
Agent plants serve as bio-sentinel nodes within the growing
environment, passively receiving VOC and acoustic signals from
stressed plants in their vicinity. The spatial coverage of the sensing
network is determined entirely by the selection of receiver positions
hosting these agent plants. If agents are clustered in one region, other
areas remain unobserved, severely degrading localization performance for
distant sources.

To avoid a data dependency between placement and evaluation,
the selection criterion must be based exclusively on field geometry,
without relying on measured signal data from the evaluation scenarios.
To satisfy this requirement, we use a geometry-driven greedy algorithm
based on QR factorization with column pivoting. This algorithm
identifies positions that maximize spatial coverage using a geometry-only
sensitivity model derived from the potential source-sensor spatial structure.
Clustering is prevented by a minimum separation constraint,
and actual source positions are excluded from the candidate set to ensure
the stressed plant itself is never designated as a sensing node.

\subsection{QR Pivoting with Minimum Separation}
A sensitivity matrix $\mathbf{S} \in \mathbb{R}^{N_c \times N_g}$ is
constructed, where $N_c$ is the number of candidate receiver positions
and $N_g$ is the number of points on a uniform source grid. Each entry
is defined as:
\begin{equation}
S_{ij} = \exp\!\left(-\frac{\|\mathbf{p}_i - \mathbf{s}_j\|^2}{2\sigma^2}\right),
\label{eq:sensitivity}
\end{equation}
where $\mathbf{p}_i$ is the position of candidate $i$, $\mathbf{s}_j$
is the $j$-th source grid point, and $\sigma = 2.0$~\si{\meter} is
the kernel width. This kernel models the spatial influence each
candidate exerts over the source grid; a large $\sigma$ promotes
broad, distributed coverage by ensuring that candidates far from the
grid center are not overly penalized.

\begin{figure*}[t!]
    \centering
    \includegraphics[height=7cm]{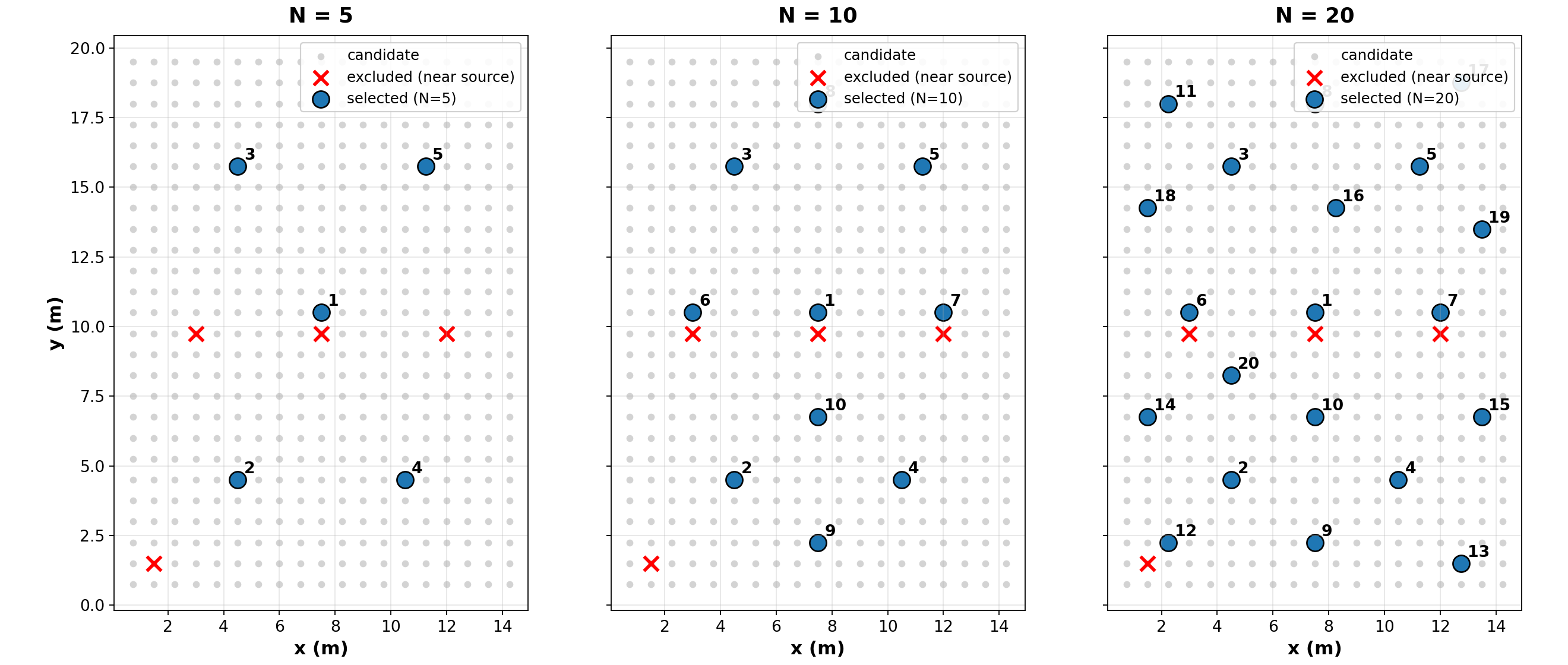}
    \caption{Greedy QR agent plant selections for $N = 5$, $10$, and
    $20$.}
    \label{fig:agent_placement}
\end{figure*}

To determine the optimal placement, QR factorization with column pivoting 
is applied to $\mathbf{S}^\top$. The pivot order ranks candidates by the 
amount of unique spatial information they contribute: each successive pivot 
selects the candidate most linearly independent from those already chosen. 
Because this greedy procedure is hierarchical, the top-$N$ selection for 
any $N$ is simply a prefix of the full ordering, ensuring all network 
sizes share a common spatial basis.

To prevent cluster formation, a minimum inter-agent separation of
1.5~\si{\meter} is enforced after each pivot selection. If the next
pivot falls within 1.5~\si{\meter} of any already-selected position,
it is skipped and the following pivot is evaluated. To
guarantee the stressed plant is not used as an agent plant, candidate positions
within 0.5~\si{\meter} of any scenario source location are excluded
prior to selection.

For these evaluations, the source grid uses 0.5~\si{\meter} spacing,
covering the full field bounding box. Network configurations are generated
for network densities of $N \in \{1, 2, 5, 10, 20, 50\}$ agent plants.

\subsection{Placement Analysis}
The resulting selections reveal a direct relationship between network
density and spatial coverage across the evaluation scenarios.

At a density of $N = 10$, the selected positions adequately cover the
central and lateral regions of the field. However, no agent is placed
near the corner source at $(1.5, 1.5)$~\si{\meter}; the nearest selected
agent remains approximately 4~\si{\meter} away. This gap is a direct
geometric consequence of the field dimensions combined with the source
exclusion constraint. Coverage of this challenging corner region is only
achieved when the density increases to $N = 20$, as illustrated in
Fig.~\ref{fig:agent_placement}. 

\section{Coarse-to-Fine Localization Framework}
\label{sec:pipeline}
This section introduces a two-stage evaluation pipeline for source
localization using acoustic (TDOA) and VOC observations. The pipeline
supports both individual-modality and combined operation, allowing TDOA-only,
VOC-only, and fused estimates to be compared directly and the relative
spatial utility of each modality to be isolated.

Stage~1 uses acoustic time-of-arrival readings from the $N$ agent plants: a
closed-form TDOA multilateration estimator produces an instantaneous bounding
region around the source in under a millisecond. Stage~2 then refines this
estimate using VOC concentration observations, via a differentiable,
physics-based inverse solver built on the closed-form Green's function of the
steady-state advection-diffusion equation, which exploits the spatial
structure of the concentration field within the bounded region.

\begin{figure*}[t]
\centering
\resizebox{\textwidth}{!}{%
\begin{tikzpicture}[
    font=\small,
    box/.style={
        draw, rectangle, rounded corners=4pt,
        minimum width=2.2cm, minimum height=0.85cm,
        text centered, align=center, line width=0.7pt
    },
    arr/.style={-{Stealth[length=4pt]}, line width=0.7pt},
    darr/.style={-{Stealth[length=4pt]}, line width=0.7pt, dashed},
    lbl/.style={font=\footnotesize, inner sep=1pt}
]

\node[box, fill=gray!10]
    (ac_in) {ToA Measurements\\($N$ agents)};
\node[box, fill=gray!10, below=1.4cm of ac_in]
    (voc_in) {VOC\\($N\!\times\!3\!\times\!120$)};

\node[box, fill=cyan!12,
      right=1.0cm of ac_in,
      minimum height=1.2cm]
    (s1) {\textbf{Stage 1}\\TDOA Localization};
\node[box, fill=gray!10,
      right=0.6cm of voc_in,
      minimum height=1.2cm]
    (vocprep) {Temporal windows\\early/mid/late\\height avg.};

\node[box, fill=cyan!6,
      right=1.0cm of s1]
    (bbox) {Bounding region\\$\hat{\mathbf{x}}_{\text{tdoa}}\!\pm\!5.5\,\mathrm{m}$};

\node[box, fill=yellow!20,
      right=2.0cm of vocprep]
    (gate) {VOC Gate\\$\mathrm{CoV} \gtrless 0.5$};

\node[box, fill=orange!15,
      right=2.2cm of bbox,
      yshift=-0.7cm,
      minimum height=2.2cm,
      minimum width=3.2cm]
    (s2) {\textbf{Stage 2 Solver}\\Steady-state Green's function\\[0.15cm] $\mathcal{L}_\mathrm{VOC}$\\[0.15cm] Adam $\cdot$ 500 ep. $\cdot$ 50 trials};

\node[box, fill=yellow!15,
      right=1.6cm of s2,
      minimum height=1.6cm,
      minimum width=2.6cm]
    (fusion) {\textbf{Inverse-Variance}\\\textbf{Fusion}\\$w_{\text{voc}} = \dfrac{1/\sigma^2_{\text{voc}}}{1/\sigma^2_{\text{voc}}+1/\sigma^2_{\text{tdoa}}}$};

\node[box, fill=green!15,
      right=1.0cm of fusion,
      minimum width=2.0cm]
    (output) {$\hat{\mathbf{x}}_{\text{fused}}$\\Source estimate};

\draw[arr] (ac_in.east)  -- (s1.west);
\draw[arr] (voc_in.east) -- (vocprep.west);

\draw[arr] (s1.east) -- (bbox.west)
    node[midway, above, lbl] {$\hat{\mathbf{x}}_{\text{tdoa}}$};

\draw[arr] (vocprep.east) -- (gate.west)
    node[midway, above, lbl] {$\mathrm{CoV}(\bar{C}_k)$};

\draw[arr] (bbox.east) -- ([yshift=0.7cm]s2.west)
    node[midway, above, lbl] {init \& clamp};

\draw[arr] (gate.east) -| ++(0.4cm,0) |- node[near start, above, lbl] {informative} ([yshift=-0.7cm]s2.west);

\draw[arr] (vocprep.south) -- ++(0,-0.3) -| (s2.south)
    node[near end, right, lbl] {obs};

\draw[darr] (gate.south) -- ++(0,-1.8) -| node[near start, below, lbl] {uninformative: bypass} (output.south);

\draw[arr] (s1.north) |- ++(0,1.2cm) -| node[near start, above, lbl] {$\hat{\mathbf{x}}_{\text{tdoa}}, \sigma^2_{\text{tdoa}}$} ([xshift=-0.6cm]fusion.north);

\draw[arr] (s2.east) -- (fusion.west)
    node[midway, above, lbl] {$\hat{\mathbf{x}}_{\text{voc}}, \sigma^2_{\text{voc}}$};

\draw[arr] (fusion.east) -- (output.west);

\end{tikzpicture}%
}
\caption{Block diagram of the coarse-to-fine localization pipeline.}
\label{fig:pipeline}
\end{figure*}
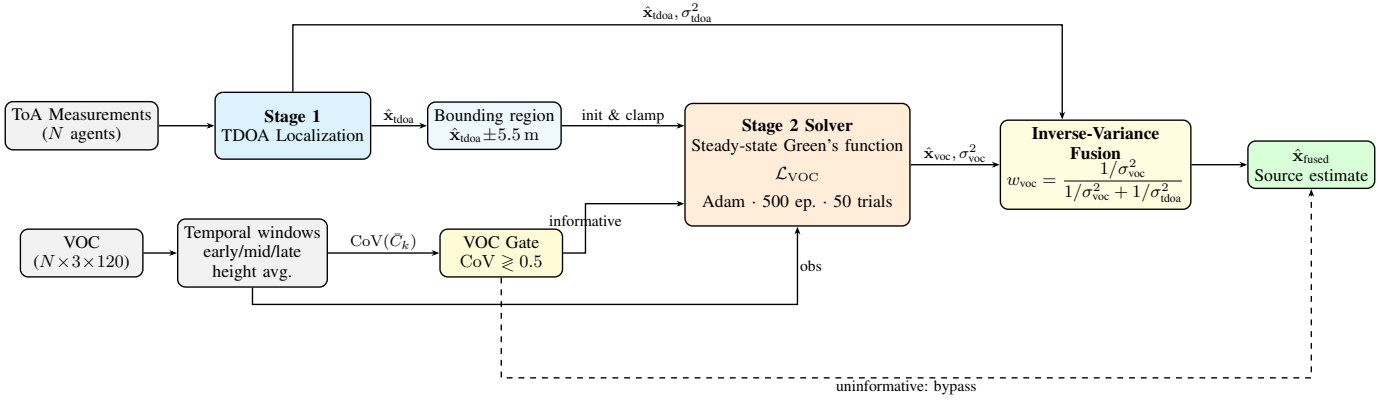

The two stages are linked by a VOC informativeness gate and an
inverse-variance fusion rule, which together transition the pipeline between
TDOA-only and VOC-informed behavior and gracefully fall back to the TDOA prior
when no informative VOC signal is available. At a glance: Stage~1 (acoustic)
yields an anchor $\hat{\mathbf{x}}_{\text{tdoa}}$ and a search region; Stage~2
(VOC) fits a source location $\hat{\mathbf{x}}_{\text{voc}}$ within that
region; the gate decides whether Stage~2's result is trusted; and fusion
combines $\hat{\mathbf{x}}_{\text{voc}}$ and $\hat{\mathbf{x}}_{\text{tdoa}}$
by their relative across-trial reliability into the final estimate
$\hat{\mathbf{x}}_{\text{fused}}$. Fig.~\ref{fig:pipeline} summarizes this
flow.

\subsection{Stage 1: Acoustic TDOA Localization}
\label{sec:stage1}
The first stage relies entirely on acoustic time-of-arrival (ToA) readings and
requires no iterative computation.

Let $\{(\mathbf{p}_k, t_k)\}_{k=1}^{N}$ denote the positions and ToA readings
of the $N$ agent plants, where $t_k$ is the propagation delay from the source
to agent $k$ along the fastest unobstructed acoustic path (direct or
ground-reflected ray), computed at $c = 343$~\si{\meter\per\second}. Only agents with an unobstructed acoustic path are used; obstructed receivers
(marked with a $-999$~\si{\second} sentinel in the dataset) are excluded. To emulate realistic clock and detection jitter, each trial draws an
independent additive white Gaussian noise (AWGN) realization on every valid
ToA, with standard deviation $\sigma_t = 0.5$~\si{\milli\second}:
\begin{equation}
\tilde{t}_k = t_k + \epsilon_k, \quad \epsilon_k \sim \mathcal{N}(0, \sigma_t^2).
\label{eq:toa_noise}
\end{equation}
If $N_v \geq 3$ agents report a valid ToA, the source location is estimated by
TDOA least squares. Choosing the earliest-arrival
agent as reference $r$, the range differences $d_k = c(\tilde{t}_k -
\tilde{t}_r)$ linearize the hyperbolic TDOA equations into
\begin{equation}
2(\mathbf{p}_k - \mathbf{p}_r)^\top \mathbf{x}_s + 2 d_k \, \rho_r
= \|\mathbf{p}_k\|^2 - \|\mathbf{p}_r\|^2 - d_k^2,
\label{eq:tdoa_lin}
\end{equation}
for every $k \neq r$, where $\rho_r = \|\mathbf{x}_s - \mathbf{p}_r\|$ is
treated as an auxiliary unknown alongside $\mathbf{x}_s \in \mathbb{R}^2$.
Stacking Eq.~\eqref{eq:tdoa_lin} over all valid $k \neq r$ yields an
overdetermined linear system $\mathbf{A}[\mathbf{x}_s^\top, \rho_r]^\top =
\mathbf{b}$, solved by ordinary least squares to give the per-trial acoustic
estimate $\hat{\mathbf{x}}_{\text{tdoa}}$, which is then clamped to the
planting area boundaries ($x \in [0.75, 14.25]$~\si{\meter}, $y \in [0.75,
19.5]$~\si{\meter}).

If $N_v < 3$, TDOA triangulation is underdetermined and the estimate falls back
to a proximity rule: the position of the agent with the earliest (noisy)
arrival, $\hat{\mathbf{x}}_{\text{tdoa}} = \mathbf{p}_{k^*}$ with $k^* =
\arg\min_{k} \tilde{t}_k$, a discrete snap to the nearest agent plant rather than a
continuous triangulated estimate. If no agent reports a valid ToA, the
estimate defaults to the domain center $(7.5, 10.0)$~\si{\meter}. Thus TDOA
triangulation tightens the acoustic-only estimate whenever three or more
agents are within range, while gracefully degrading to the proximity rule in
sparse networks where triangulation is not geometrically well-posed.

Each of the $50$ trials draws an independent noise realization
$\{\epsilon_k\}$, producing a per-trial anchor
$\hat{\mathbf{x}}_{\text{tdoa}}^{(t)}$ and a corresponding bounding region of
radius $r = 5.5$~\si{\meter} centered on it (room-clamped; trial~0 is
initialized at its anchor, trials $1$--$49$ drawn uniformly within the box).
This radius corresponds approximately to one agent-spacing at $N = 10$ and is
held fixed across network densities; its sensitivity, and that of $\sigma_t$,
are evaluated in subsequent sections. At
$\sigma_t = 0$, all trials share one anchor and the pipeline reduces to a
single noiseless TDOA estimate with multi-start initialization.

\subsection{Stage 2: VOC Physics-Based Inverse Solver with a Steady-State Advection-Diffusion Green's Function}
\label{sec:stage2}
Within the bounding region established by Stage~1, the inverse solver
refines the source estimate by leveraging the spatial structure of the VOC
observations. This forward model is an intentional surrogate rather than a
rigorous inverse of the dataset generator: the dataset is produced by a full
3D finite-volume solver with open boundaries, no-flux surfaces, and a
transient release, whereas the Green's-function model below assumes a
steady-state, free-field point source with an effective diffusivity decoupled
from $D_\text{VOC}$. This trade-off enables a closed-form, differentiable
model applicable to any deployment without geometry-specific
re-parameterization; the robustness analysis in Section~\ref{sec:eval}
confirms the mismatch does not produce significant localization error.

The source coordinate $\mathbf{x}_s \in \mathbb{R}^2$ serves as the sole
learnable parameter. VOC observations are extracted from the dataset by
computing the mean concentration over three temporal windows---early
($t = 1$--$20$~\si{\second}), mid ($t = 41$--$80$~\si{\second}), and late
($t = 101$--$120$~\si{\second})---at each agent plant. This yields an
observation tensor of shape $(N, 3, 3)$ representing agents, heights, and
temporal stages. Observations are then averaged across heights and across the
three temporal stages, producing a single time-averaged receiver-mean
concentration $\bar{C}_k$ per agent. The vector $\{\bar{C}_k\}_{k=1}^N$ is then
normalized to sum to unity across agents, removing any dependence on absolute
emission rates and yielding the fitting target $\tilde{C}_k$.

The forward model predicts the steady-state concentration at agent $k$ using
the closed-form Green's function of the 2D advection-diffusion equation for a
continuous point source anchored at $\mathbf{x}_s$:
\begin{equation}
\hat{C}_k \propto \exp\!\left(\frac{|\mathbf{u}|\, \delta_{\parallel,k}}{2D}\right)
K_0\!\left(\frac{|\mathbf{u}|\, r_k}{2D}\right),
\label{eq:greens}
\end{equation}
where $\boldsymbol{\delta}_k = \mathbf{p}_k - \mathbf{x}_s$ is the displacement
from the candidate source to agent $k$, $r_k = \|\boldsymbol{\delta}_k\|$,
$\delta_{\parallel,k} = \boldsymbol{\delta}_k \cdot \hat{\mathbf{u}}$ is its
along-wind component, $\hat{\mathbf{u}} = \mathbf{u}/|\mathbf{u}|$ is the unit
wind direction, and $K_0$ is the modified Bessel function of the second kind
of order zero. The exponential prefactor encodes the upwind/downwind asymmetry
of an advected plume, while $K_0$ governs the radial decay of the steady-state
field; together they reproduce the characteristic narrow, wind-aligned plume
shape without requiring an explicit time-stepped simulation.

When $|\mathbf{u}| < 10^{-3}$~\si{\meter\per\second} (effectively still air),
Eq.~\eqref{eq:greens} is replaced by its diffusion-only limit, an isotropic
Gaussian:
\begin{equation}
\hat{C}_k \propto \exp\!\left(-\frac{r_k^2}{2(2D\,T + \sigma_f^2)}\right),
\label{eq:greens_diffusion}
\end{equation}
where $T = 120$~\si{\second} is the simulation duration and $\sigma_f =
0.5$~\si{\meter} is a sub-grid mixing floor.

The diffusivity $D = 0.1$~\si{\meter\squared\per\second} is an effective
turbulent value (not the molecular $D_\text{VOC}$ from
Table~\ref{tab:simparams}), chosen so the Green's-function plume half-width
$2D/|\mathbf{u}| \approx 0.4$~\si{\meter} at $|\mathbf{u}| =
0.5$~\si{\meter\per\second} is on the order of the inter-plant spacing and
consistent with reported eddy diffusivities for canopy-scale scalar transport
($10^{-2}$--$10^{-1}$~\si{\meter\squared\per\second}). The emission rate $Q_v$
is held at an arbitrary constant, since the loss is computed on
shape-normalized concentrations and is invariant to absolute scale. For
numerical stability, $r_k$ is floored as $r_k \rightarrow \sqrt{r_k^2 +
r_{\min}^2}$ with $r_{\min} = 0.05$~\si{\meter} to regularize $K_0$ near the
source, and both the exponential prefactor and $K_0$ are evaluated in the log
domain with per-trial max-subtraction before normalization.

For each trial $t$, the predicted field is shape-normalized,
$\tilde{\hat{C}}_k^{(t)} = \hat{C}_k^{(t)} / \sum_j \hat{C}_j^{(t)}$, and the
VOC loss accumulated over all trials is:
\begin{equation}
\mathcal{L}_{\text{VOC}} = \sum_{t=1}^{50}
\frac{1}{N}\left\| \tilde{C} - \tilde{\hat{C}}^{(t)} \right\|^2,
\label{eq:voc_loss}
\end{equation}
where the $1/N$ factor normalizes the loss by the number of agent plants,
ensuring the VOC loss magnitude remains independent of network density. Each
trial $t$ is optimized independently via Adam for 500 epochs at a learning
rate of 0.05, with $\mathbf{x}_s$ clamped to the Stage~1 bounding region after
every gradient step, yielding a per-trial converged estimate
$\hat{\mathbf{x}}_{\text{voc}}^{(t)}$.

\subsection{VOC Informativeness Gate}
\label{sec:gate}
Not every scenario provides a VOC signal that is informative enough to
support source localization. When ambient turbulence disperses the plume
nearly uniformly across the deployment area, or when the source is far from
all agents, the VOC readings carry little spatial structure and optimizing
Eq.~\eqref{eq:voc_loss} can pull the estimate away from the true source. To
guard against this, we introduce an informativeness gate based on the
coefficient of variation of the raw agent-mean VOC readings,
\begin{equation}
\mathrm{CoV} = \frac{\mathrm{std}(\bar{C}_1,\dots,\bar{C}_N)}
                     {\mathrm{mean}(\bar{C}_1,\dots,\bar{C}_N)}.
\label{eq:voc_cov}
\end{equation}
If $\mathrm{CoV} \geq 0.5$, the VOC readings are considered sufficiently
informative and Stage~2 proceeds as described above. If
$\mathrm{CoV} < 0.5$, the VOC signal is treated as uninformative, Stage~2 is
bypassed entirely, and the fused estimate (Section~\ref{sec:fusion}) reduces
to the TDOA estimate $\hat{\mathbf{x}}_{\text{tdoa}}^{(t)}$ for every trial ---
except when Stage~1 itself is in proximity fallback, where the TDOA anchor is unreliable and
Stage~2 proceeds regardless of CoV, since VOC is the only remaining source of
spatial information.

\subsection{Inverse-Variance Multimodal Fusion}
\label{sec:fusion}
When the VOC gate is open, the per-trial VOC and TDOA estimates are combined
using an inverse-variance (precision-weighted) fusion rule, which assigns
greater weight to whichever modality exhibits lower across-trial scatter. The
VOC and TDOA variances are each taken as the mean per-coordinate variance of
their fifty per-trial solutions, with floors that prevent division by zero or
unrealistic confidence in near-degenerate clusters:
\begin{equation}
\sigma^2_{\text{voc}} = \max\!\left(
\mathrm{Var}\big(\hat{\mathbf{x}}_{\text{voc}}^{(1)},\dots,\hat{\mathbf{x}}_{\text{voc}}^{(50)}\big),\;
1.0~\si{\meter\squared}\right),
\label{eq:sigma_voc}
\end{equation}
\begin{equation}
\sigma^2_{\text{tdoa}} = \max\!\left(
\mathrm{Var}\big(\hat{\mathbf{x}}_{\text{tdoa}}^{(1)},\dots,\hat{\mathbf{x}}_{\text{tdoa}}^{(50)}\big),\;
10^{-6}~\si{\meter\squared}\right).
\label{eq:sigma_tdoa}
\end{equation}
If fewer than three agents report arrival times and Stage~1 falls back to the
proximity estimate (Section~\ref{sec:stage1}), the TDOA solution carries no
continuous spread, so $\sigma^2_{\text{tdoa}}$ is instead fixed to a large
fallback value reflecting the reduced confidence in a single-anchor estimate,
\begin{equation}
\sigma^2_{\text{tdoa}} = 25~\si{\meter\squared}.
\label{eq:sigma_tdoa_fallback}
\end{equation}
The fusion weight assigned to the VOC estimate is then
\begin{equation}
w_{\text{voc}} = \frac{1/\sigma^2_{\text{voc}}}
{1/\sigma^2_{\text{voc}} + 1/\sigma^2_{\text{tdoa}}},
\label{eq:fusion_weight}
\end{equation}
and the fused per-trial source estimate is
\begin{equation}
\hat{\mathbf{x}}_{\text{fused}}^{(t)} =
w_{\text{voc}}\, \hat{\mathbf{x}}_{\text{voc}}^{(t)}
+ (1 - w_{\text{voc}})\, \hat{\mathbf{x}}_{\text{tdoa}}^{(t)}.
\label{eq:fused_estimate}
\end{equation}
Intuitively, $w_{\text{voc}} \to 1$ when the VOC solutions are tightly
clustered relative to TDOA, so the fused estimate is dominated by the VOC
inversion; conversely $w_{\text{voc}} \to 0$ when TDOA is tightly clustered
(e.g., a well-conditioned multilateration geometry) and VOC is scattered, so
the estimate reverts toward the TDOA anchor. This data-driven weighting
adapts per scenario without a hand-tuned constant.
\section{Results and Analysis}
\label{sec:eval}
This section evaluates TDOA, VOC, and multimodal fusion across agent plant
availability, parameter and sensor robustness, and agent placement. The
pipeline of Section~\ref{sec:pipeline} allows each stage to run independently
under identical dataset and placement conditions: Stage~1 alone gives the
TDOA-only result; Stage~2 alone---initialized with a room-center prior in
place of the TDOA anchor---gives the VOC-only result; and the gated,
inverse-variance-fused combination gives the multimodal result. Because
TDOA-only bypasses Stage~2 entirely, it is unaffected by the physical
transport model; the parameter-perturbation analysis therefore concerns only
the VOC and fusion variants, while the sensor-threshold analysis evaluates
each modality independently. We frame this as a modality comparison rather
than a benchmark against prior localization algorithms, as such algorithms
are not directly applicable here: existing acoustic and chemical-plume
localization methods assume dedicated, actively-placed sensor arrays, whereas
the IoP setting senses both modalities passively through living organisms at
fixed agricultural positions, with no comparable prior baseline to adapt.

All results report the mean absolute error (MAE) over 50 trials per scenario
across the full 52-scenario dataset, together with the median, 90th-percentile
(P90), and success rates within 0.75~\si{\meter} and 1.0~\si{\meter}
($\mathrm{SR}_{0.75}$, $\mathrm{SR}_{1.0}$). The availability analysis covers
$N \in \{1, 2, 5, 10, 20, 50\}$; all other analyses use $N \in \{5, 10, 20\}$.

\subsection{Modality Ablation}
\label{sec:ablation}
We evaluated the three pipeline variants---TDOA-only, VOC-only, and multimodal fusion---
across all 52 scenarios, grouped by source position to isolate the
contribution of each modality.

\begin{figure}[t]
    \centering
    \includegraphics[width=\linewidth]{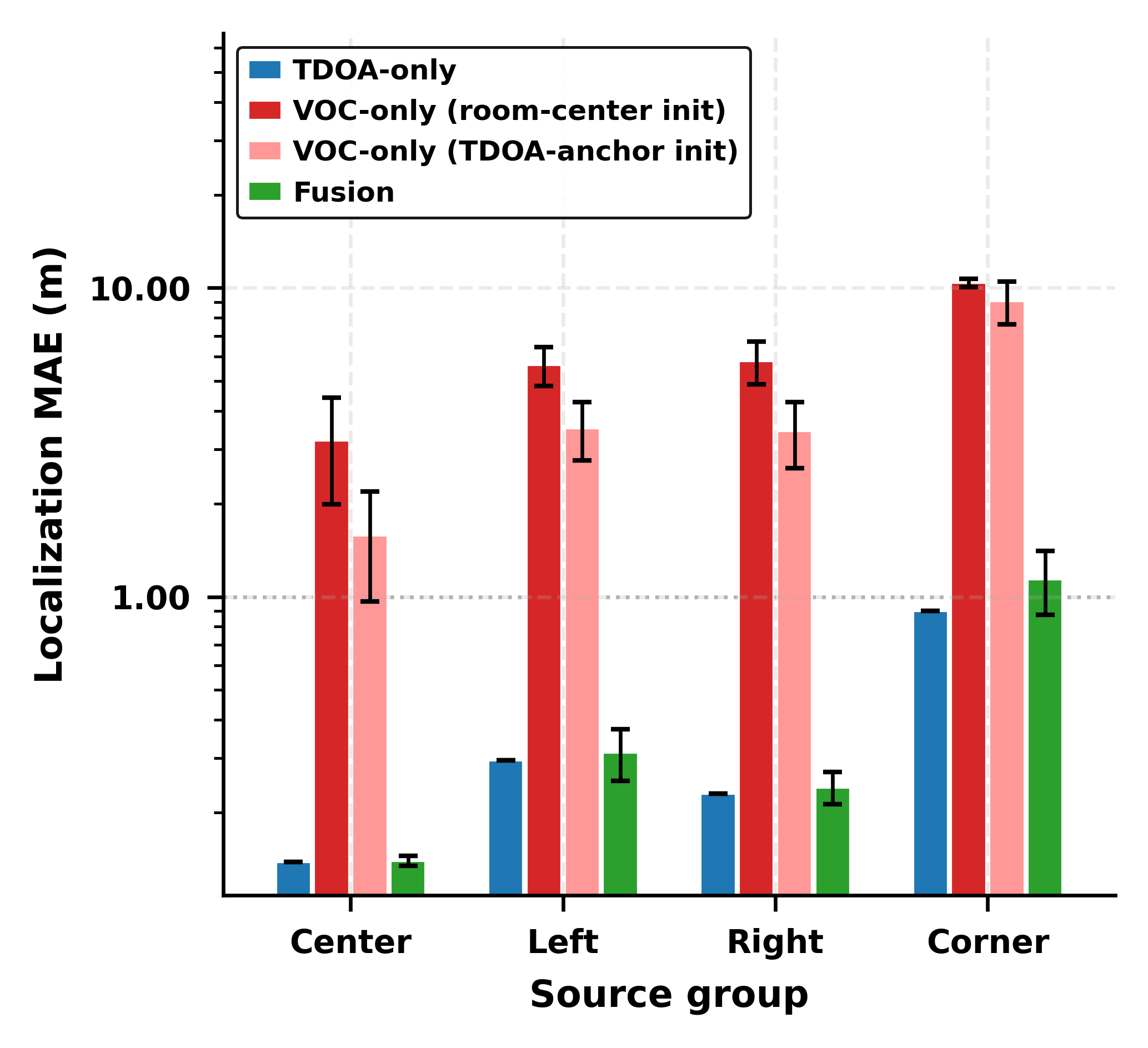}
    \caption{Localization MAE (m, mean $\pm$ 95\,\% CI across scenarios in
    each group, log scale) per modality and source group.}
    \label{fig:ablation_modality}
\end{figure}

At $N = 10$ (Fig.~\ref{fig:ablation_modality}), TDOA-only localizes the
central source to $0.14$~\si{\meter} and the lateral sources to
$0.23$--$0.30$~\si{\meter}, at or below one grid spacing. VOC-only
(room-center initialization) is roughly an order of magnitude worse, at
$3.20$~\si{\meter} for the central source and $5.63$--$10.40$~\si{\meter} for
the lateral and corner sources, and remains above $3$~\si{\meter} for every
source group at every density tested. This stems from the geometry of the
problem: the VOC plume is narrow and advection-dominated, so most agents
never observe a useful concentration signal regardless of network density.

This large gap raises a natural question: is VOC-only weak because of the
forward model itself, or because it starts from a poor initial guess (the
room center)? To isolate these factors, we re-ran VOC-only with Stage~2
initialized at the TDOA anchor rather than the room center, optimizing using
only the VOC loss thereafter. This substantially improves performance (e.g.,
the central-source error drops from $3.20$ to $1.58$~\si{\meter} at $N=10$),
but VOC-only still falls well short of TDOA-only or fusion. Initialization
therefore accounts for most---but not all---of VOC's weakness: even given a
favorable starting point, the VOC loss surface is too flat and multi-modal
away from the plume centerline to permit substantial further refinement.
Accordingly, we treat VOC throughout this paper as a local refinement
signal rather than a standalone localizer, a framing that also accounts for
the following two results.

First, it explains an apparent anomaly at $N=5$. There, TDOA triangulation is
underdetermined for the corner source (fewer than three agents report a valid
ToA reading), so Stage~1 falls back to a coarse single-anchor rule and
produces a $14.56$~\si{\meter} outlier. This single case raises the overall
$N=5$ mean to $3.88$~\si{\meter} for TDOA (versus $5.92$~\si{\meter} for VOC),
even though TDOA still clearly outperforms VOC for the other three source
groups ($0.26$--$0.43$~\si{\meter}). By $N=20$, enough agents fall within
range of every source position and TDOA again leads across the board,
including the corner ($0.68$~\si{\meter}). As above, this is a density effect
on TDOA rather than evidence that VOC becomes competitive.

Second, it establishes expectations for fusion. Because VOC-only remains far
behind TDOA-only, fusion has limited room to improve on it at present: across
most scenarios, the fused estimate is statistically indistinguishable from
the better single-modality result, with differences typically under
$0.02$~\si{\meter}---within the 95\% CI, and therefore attributable to noise.
The one partial exception is the corner-source group, the same case discussed
above, where fusion does not reduce the proximity-fallback outlier relative to
TDOA-only ($0.90$ and $0.68$~\si{\meter} at $N=10$ and $N=20$, versus
$1.14$ and $0.77$~\si{\meter} for fusion). Here the gate correctly limits
VOC's contribution, keeping the fused estimate close to the TDOA-only baseline
rather than degrading further toward the VOC-only error ($9.06$ and
$5.88$~\si{\meter}). The more
significant point, however, is structural rather than numerical: because the
gate and inverse-variance weighting scale $w_{\text{voc}}$ according to the
informativeness of the VOC signal, the fusion mechanism is positioned to
benefit automatically from future improvements in VOC sensing and modeling,
without requiring changes to the pipeline itself.

\subsection{Effect of Agent Plant Availability}
\label{sec:availability}
This subsection examines how the three pipeline variants---TDOA-only,
VOC-only, and multimodal fusion---respond to network density, by evaluating
all 52 scenarios at $N \in \{1, 2, 5, 10, 20, 50\}$ and tracking the overall
mean MAE.

\begin{figure}[t]
    \centering
    \includegraphics[width=\linewidth]{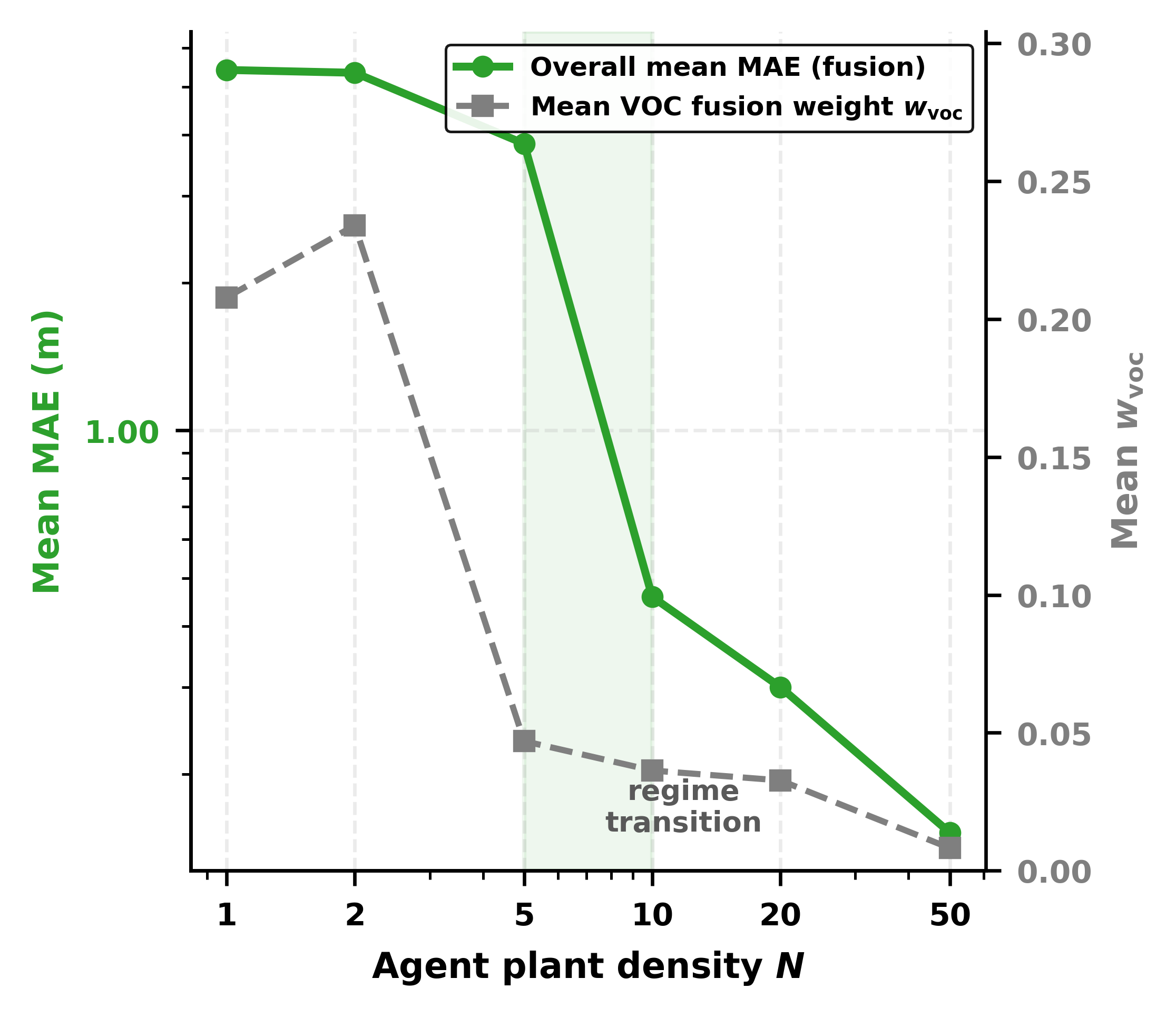}
    \caption{Fusion pipeline mean localization MAE (left axis, log scale) and
    mean VOC fusion weight $w_{\text{voc}}$ (right axis) across agent plant
    densities $N \in \{1, 2, 5, 10, 20, 50\}$.}
    \label{fig:availability}
\end{figure}

Fig.~\ref{fig:availability} shows a clear regime transition between $N=5$ and
$N=10$. Below this point, TDOA multilateration is poorly conditioned for at
least one source group---most notably the corner source, whose
proximity-fallback outlier was discussed in
Section~\ref{sec:ablation}---and the overall mean MAE remains high, at
$3.83$~\si{\meter} for $N=5$. Once $N \geq 10$, at least three agents are
audible for every source group, multilateration becomes well-posed
throughout, and the overall mean MAE drops sharply to $0.46$~\si{\meter}
(median $3.78 \to 0.43$~\si{\meter}, $\mathrm{SR}_{0.75}$ $0.71 \to 0.81$). The
trend continues smoothly with further increases in density, reaching
$0.15$~\si{\meter} (median $0.14$, $\mathrm{SR}_{0.75}=0.99$,
$\mathrm{SR}_{1.0}=1.00$) by $N=50$.

This transition is driven entirely by the acoustic modality: as density
increases, more agents fall within range of each source, multilateration
becomes increasingly well-conditioned, and the TDOA estimate improves
correspondingly. VOC does not contribute to this trend in the same way---its
accuracy remains limited by plume geometry rather than agent count
(Section~\ref{sec:ablation})---but the mean fusion weight $w_{\text{voc}}$,
also shown in Fig.~\ref{fig:availability}, reflects this asymmetry directly.
At very low densities ($N=1,2$), where TDOA itself is poorly conditioned and
offers little advantage over VOC, $w_{\text{voc}}$ is comparatively high
($0.21$ and $0.23$, respectively). As TDOA solutions become tightly clustered
with increasing density, the gate and inverse-variance rule correspondingly
suppress VOC, and $w_{\text{voc}}$ falls to $0.05$ or below for $N \geq 5$.

\begin{figure}[t]
    \centering
    \includegraphics[width=\linewidth]{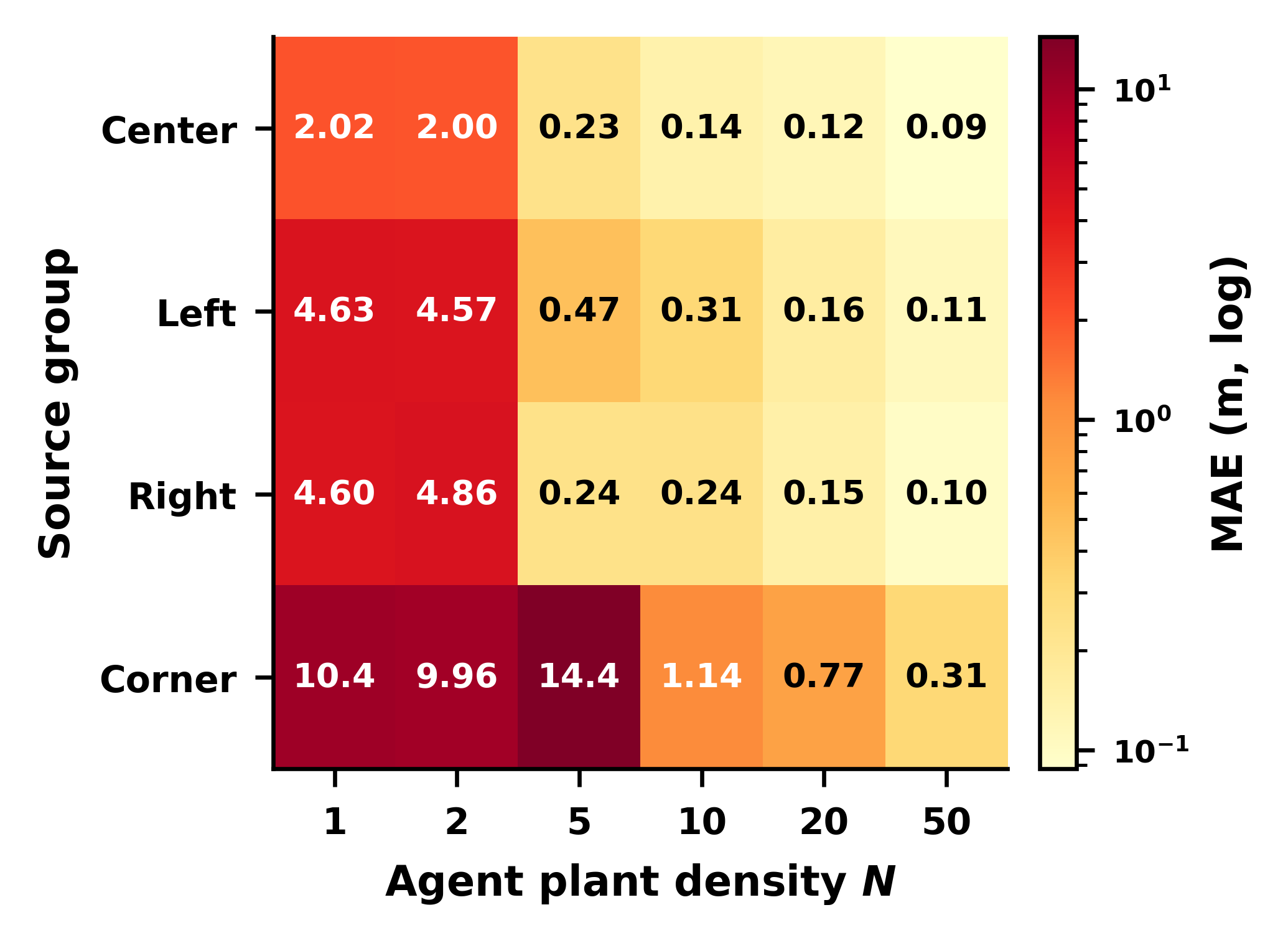}
    \caption{Fusion pipeline mean MAE (m) per source group (Center, Left, Right,
    Corner) across network densities $N \in \{1, 2, 5, 10, 20, 50\}$, shown as a
    heatmap.}
    \label{fig:persource}
\end{figure}

Taken together, these results show that network density governs localization
accuracy primarily through the acoustic modality: increasing $N$ improves the
conditioning of TDOA multilateration, which drives the overall MAE down and,
in turn, reduces the role of VOC in the fused estimate.

Breaking this trend down by source position (Fig.~\ref{fig:persource})
confirms that it holds throughout the domain, but also reveals a persistent
asymmetry. The Center, Left, and Right groups follow a similar trajectory:
large errors at $N \in \{1,2\}$ ($2.0$, and $4.6$--$4.9$~\si{\meter},
respectively), a sharp drop once $N \geq 5$ brings at least three agents into
range, and a gradual decline to $0.09$--$0.11$~\si{\meter} at $N=50$. The
Corner group follows the same pattern but lags by roughly one density tier:
above $9.9$~\si{\meter} for $N \in \{1,2\}$, only partially improved at $N=5$
($14.40$~\si{\meter}, the proximity-fallback effect discussed above), then
$1.14$, $0.77$, and $0.31$~\si{\meter} at $N=10$, $20$, and $50$---still
$2$--$3\times$ the other groups at the same density. This persistent gap
reflects the corner source's position at the domain boundary
$(1.5,1.5)$~\si{\meter}, where roughly a quarter of the surrounding area lies
outside the planting domain, leaving fewer agents in range for any given $N$.

For the area studied, $N \geq 5$ is
therefore generally sufficient for well-posed triangulation of central and
lateral sources, while the corner source---the binding constraint on minimum
viable network density---requires $N \geq 10$.

\subsection{Robustness and Sensitivity Analyses}
\label{sec:robustness}
The following three subsections evaluate the pipeline's sensitivity to
physical parameter uncertainty, sensor detection floors, and the principal
hyperparameters introduced in Section~\ref{sec:pipeline}: the Stage~1 bounding
radius $r$, the ToA noise and bias levels $\sigma_t$ and $\sigma_b$, and the
VOC gate threshold $\tau$ (Section~\ref{sec:gate}).

\subsubsection{Robustness to Parameter Uncertainty}
\begin{figure}[t]
\centering
\includegraphics[width=\linewidth]{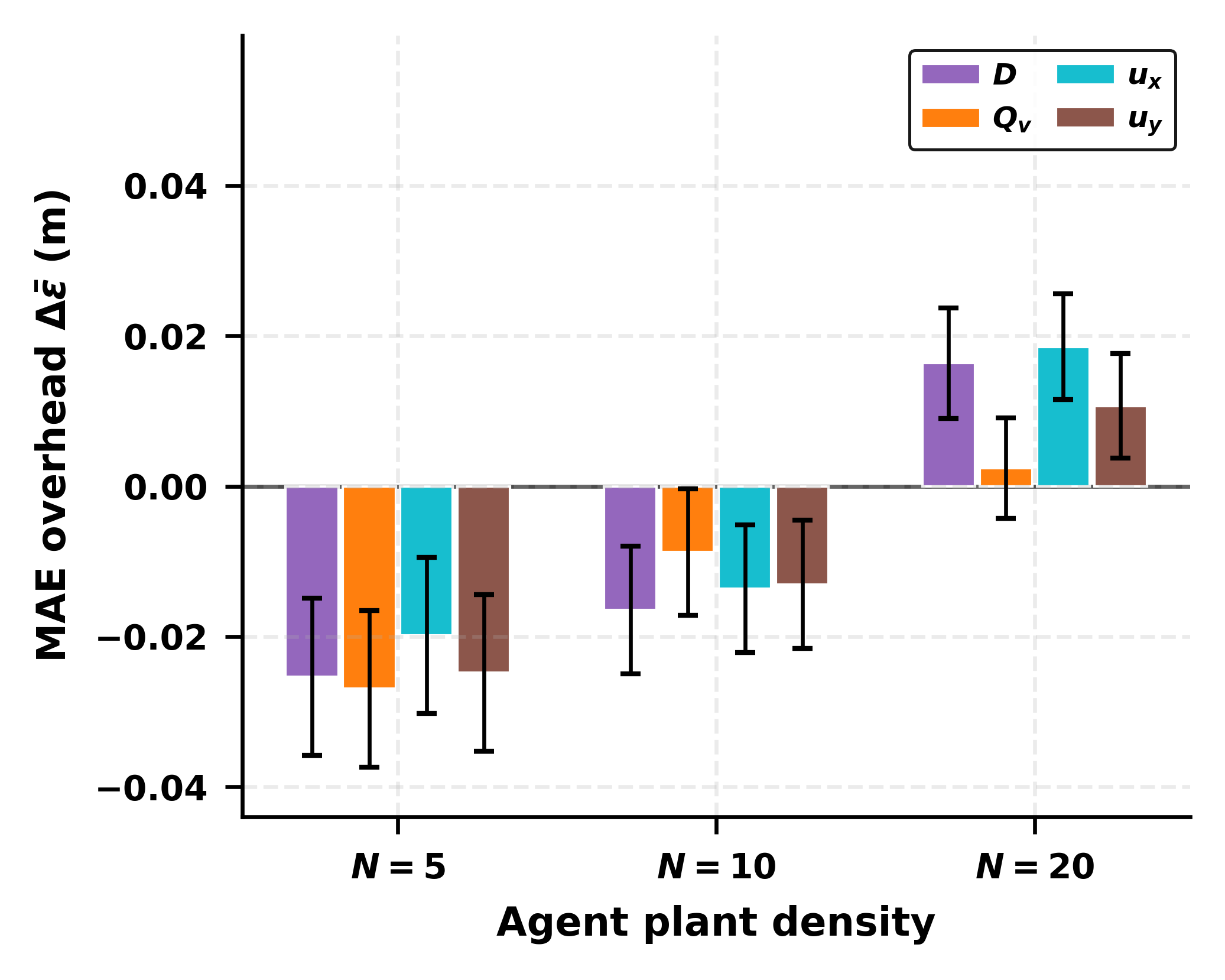}
\caption{Fusion MAE overhead $\Delta\bar{\varepsilon}$ (m, mean $\pm$ 95\,\%
CI of the perturbed condition) under 40\,\% Gaussian perturbation of each
physical parameter, relative to baseline, at $N \in \{5, 10, 20\}$.}
\label{fig:perturbation}
\end{figure}

Field deployments rarely know the physical parameters governing VOC transport
precisely. We applied one-at-a-time 40\% Gaussian perturbations
($\mathcal{N}(1.0, 0.4)$) to the diffusivity $D$, the
emission intensity $Q_v$, and the wind components $u_x$ and $u_y$, with the
full multi-start fusion pipeline (50 trials) used for both baseline and
perturbed conditions. Fig.~\ref{fig:perturbation} reports the resulting MAE
overhead relative to the baseline fusion MAE ($3.83 \pm 0.24$, $0.46 \pm
0.02$, and $0.30 \pm 0.02$~\si{\meter} at $N=5$, $10$, and $20$).

The overhead never exceeds $0.03$~\si{\meter} in magnitude at any density and
is in every case smaller than the 95\% CI of the perturbed mean---
statistically indistinguishable from zero. Two mechanisms explain this: the
VOC informativeness gate (Section~\ref{sec:gate}) bypasses Stage~2 entirely
when VOC signals are weak, reducing the fused estimate to the TDOA solution
regardless of transport parameters; and when the gate is open, the
inverse-variance rule (Section~\ref{sec:fusion}) automatically lowers
$w_{\text{voc}}$ for any parameter perturbation that increases VOC solution
scatter. The TDOA stage and gating/fusion mechanism together act as a stable
performance floor, so parameter uncertainty in the VOC transport model does
not propagate into localization error in any practically significant way.

\subsubsection{Sensor Threshold Sensitivity}
\label{sec:thresholds}
Real-world sensors have a detection floor below which readings are
indistinguishable from noise. We applied thresholds at the 25th, 50th, 75th,
and 95th percentiles of each modality's observed signal distribution at
$N=10$, evaluating each modality independently: acoustic ToA thresholds via
the TDOA-only pipeline, and VOC thresholds via the VOC-only pipeline. A
$P_{95}$ tier retains only the top 5\% of readings by signal strength,
simulating an exceptionally high detection floor.

\begin{figure}[t]
\centering
\includegraphics[width=\linewidth]{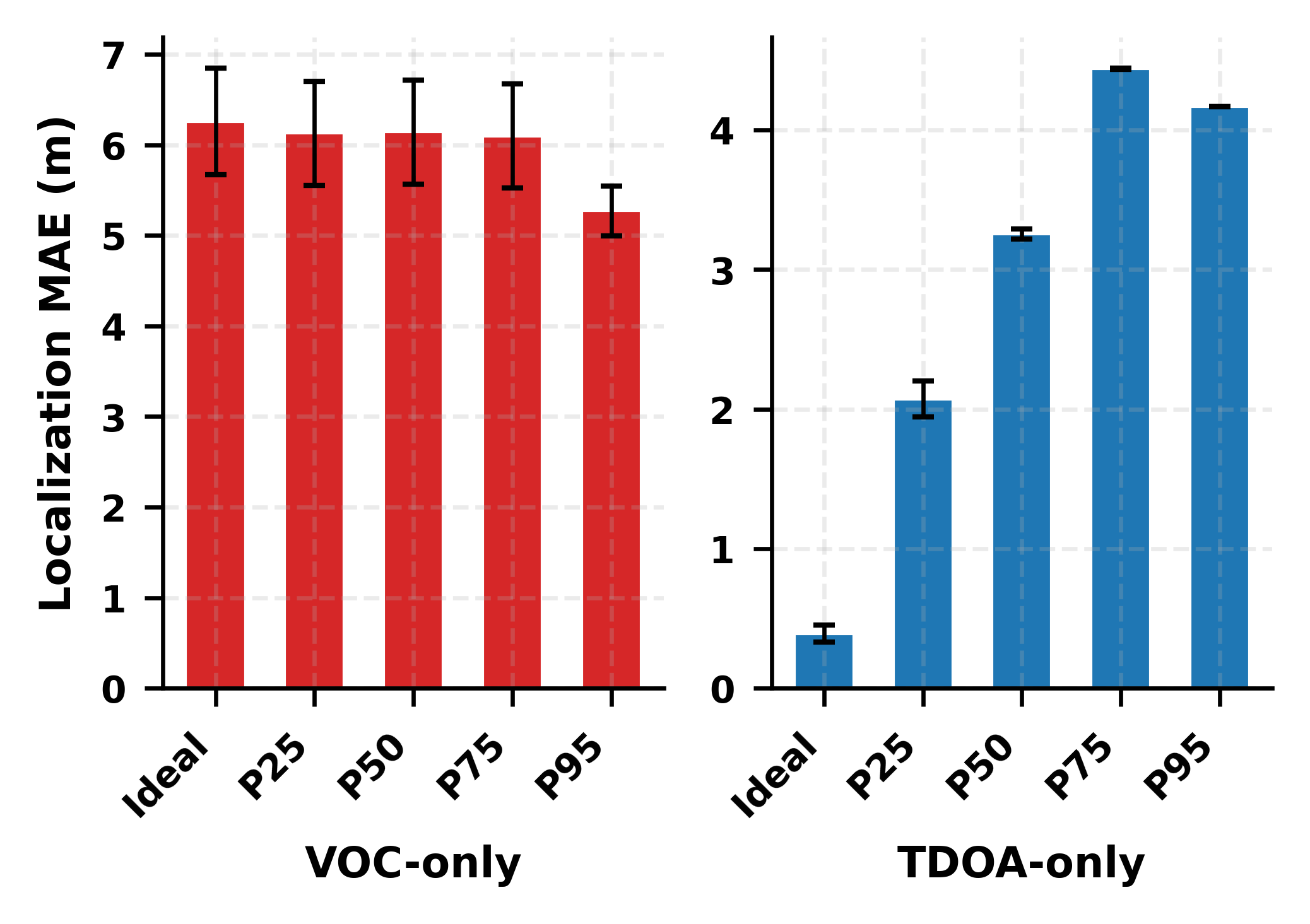}
\caption{VOC-only and TDOA-only localization MAE (m, mean $\pm$ 95\% CI
pooled across all 52 evaluated scenarios) across sensor threshold tiers
($N = 10$).}
\label{fig:thresholds}
\end{figure}

VOC localization is highly insensitive to thresholding
(Fig.~\ref{fig:thresholds}, left): MAE stays essentially flat at
$6.10$--$6.26$~\si{\meter}, improving only mildly to $5.27$~\si{\meter} at the
extreme $P_{95}$ tier. Because the advection-dominated plume is so narrow,
most agents report near-zero concentrations regardless of threshold, so
gating out negligible readings changes neither the typical outcome nor its
spread---performance is limited by plume-agent geometry, not sensor
sensitivity. Peak simulated VOC concentrations across all scenarios reached
approximately $0.05$~ppb (${\approx}2$~\si{\nmol\per\meter\cubed}), below the
detection limit of current IoT-grade chemical sensors~\cite{Ibrahim2022ACS,
Lee2024AdvSustain}.

TDOA localization, by contrast, is highly sensitive to thresholding
(Fig.~\ref{fig:thresholds}, right). The ideal, unthresholded MAE of
$0.39$~\si{\meter} degrades sharply once readings are gated: $2.07$~\si{\meter}
at $P_{25}$ ($7.8$~\si{\decibel SPL}), $3.26$~\si{\meter} at $P_{50}$
($19.4$~\si{\decibel SPL}), and $4.44$~\si{\meter} at $P_{75}$
($25.3$~\si{\decibel SPL}), plateauing at $4.17$~\si{\meter} at $P_{95}$
($46.5$~\si{\decibel SPL}). This reflects the loss of multilateration
geometry: as weaker, more distant readings are discarded, the number of valid
readings $N_v$ frequently drops below the three required for a well-posed
TDOA solution, forcing an increasing share of scenarios onto the single-anchor
proximity rule (Section~\ref{sec:stage1}); at $P_{95}$, nearly all scenarios
have fallen back, so the MAE plateaus near the proximity-rule error rather
than continuing to grow. In short, TDOA's accuracy gain over single-anchor
proximity depends on enough agents clearing a modest detection floor (below
approximately $20$~\si{\decibel SPL}); above this floor it degrades toward,
but not below, the proximity-rule baseline.

Together, these results indicate TDOA localization is actionable with today's
hardware up to a moderate detection floor (below approximately
$20$~\si{\decibel SPL}), while VOC localization remains forward-looking,
dependent on ultra-sensitive sub-ppb chemical sensors and higher network
densities.

\subsubsection{Hyperparameter Robustness}
\label{sec:hyperparams}
The pipeline introduces four further hyperparameters, each swept in turn with
the full fusion pipeline at $N \in \{5, 10, 20\}$: the Stage~1 bounding radius
$r \in \{2.0, 3.5, 5.5, 7.5, 10.0, 15.0\}$~\si{\meter} (default $5.5$, chosen to
represent approximately one agent-spacing at $N=10$); the ToA jitter $\sigma_t
\in \{0, 0.1, 0.3, 0.5, 1, 3, 10\}$~\si{\milli\second} (default $0.5$); a fixed
per-agent clock bias $\sigma_b \in \{0, 0.5, 1, 2, 3\}$~\si{\milli\second},
applied on top of the default $\sigma_t$ per $\tilde{t}_k = t_k + b_k +
\epsilon_k$ with $b_k \sim \mathcal{N}(0, \sigma_b^2)$; and the VOC gate
threshold $\tau \in \{0, 0.1, 0.25, 0.5, 1.0, 2.0, 5.0\}$ (default $0.5$,
Section~\ref{sec:gate}).

\begin{figure}[t]
\centering
\includegraphics[width=\linewidth]{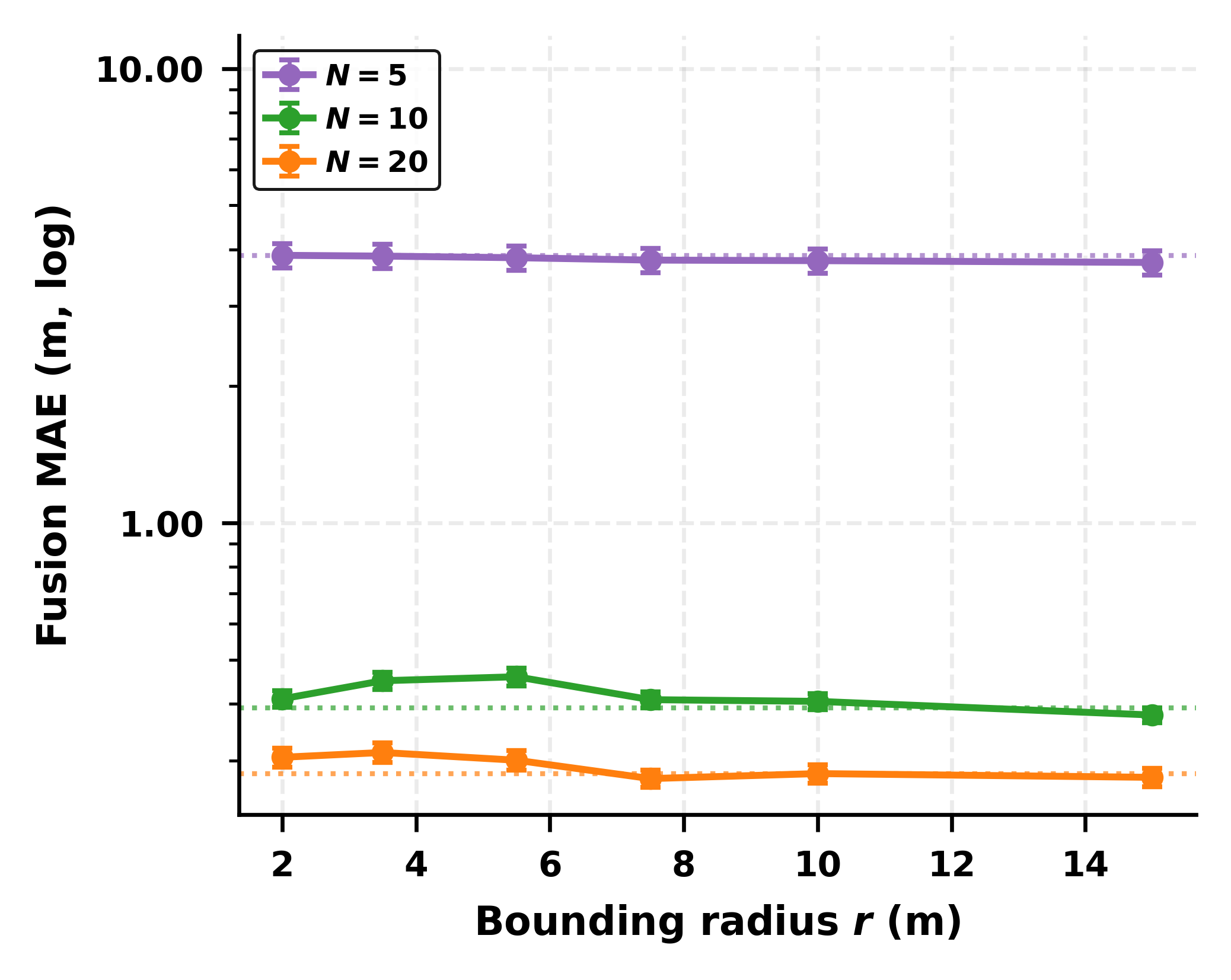}
\caption{Fusion pipeline MAE (m, mean $\pm$ 95\% CI, log scale) as a
function of bounding radius $r$ at $N \in \{5, 10, 20\}$. Dotted lines
mark the radius-independent TDOA-only reference for each density.}
\label{fig:radius}
\end{figure}

\begin{table}[b]
\centering
\caption{Fusion MAE (m, mean $\pm$ 95\% CI) across the VOC gate threshold
$\tau$ sweep. The default $\tau = 0.5$ is underlined.}
\label{tab:gate}
\footnotesize
\begin{tabular}{c ccc}
\toprule
$\tau$ & $N{=}5$ & $N{=}10$ & $N{=}20$ \\
\midrule
$0$    & $3.82{\pm}0.23$ & $0.43{\pm}0.02$ & $0.30{\pm}0.01$ \\
$0.1$  & $3.80{\pm}0.23$ & $0.45{\pm}0.02$ & $0.31{\pm}0.02$ \\
$0.25$ & $3.81{\pm}0.23$ & $0.48{\pm}0.02$ & $0.29{\pm}0.01$ \\
$\underline{0.5}$  & $\underline{3.83{\pm}0.24}$ & $\underline{0.46{\pm}0.02}$ & $\underline{0.30{\pm}0.02}$ \\
$1.0$  & $3.82{\pm}0.24$ & $0.45{\pm}0.02$ & $0.32{\pm}0.02$ \\
$2.0$  & $3.80{\pm}0.23$ & $0.45{\pm}0.02$ & $0.29{\pm}0.01$ \\
$5.0$  & $3.80{\pm}0.23$ & $0.38{\pm}0.01$ & $0.27{\pm}0.01$ \\
\bottomrule
\end{tabular}
\end{table}

Bounding radius $r$ and gate threshold $\tau$ are both effectively inert.
Fusion MAE does not trend monotonically with $r$
(Fig.~\ref{fig:radius}): at $N=10$ it varies only between $0.38$ and
$0.46$~\si{\meter} (TDOA-only reference $0.39$), and at $N=20$ between $0.27$
and $0.31$~\si{\meter} (reference $0.28$), staying within a few centimeters of
the TDOA reference regardless of $r$ because the gate and inverse-variance
fusion (Section~\ref{sec:fusion}) prevent the VOC solution from perturbing an
already-precise anchor even when the bounding region is large. Similarly,
Table~\ref{tab:gate} shows the fused MAE varies by no more than
$0.10$~\si{\meter} across the full $\tau$ sweep at any density: at $\tau=0$
(Stage~2 always runs) the inverse-variance rule already suppresses
uninformative VOC solutions via $\sigma^2_{\text{voc}}$
(Eq.~\eqref{eq:fusion_weight}) even without the gate discarding Stage~2
outright, while at $\tau=5.0$ (most restrictive) the estimate simply reduces
toward TDOA-only. The default values $r=5.5$ and $\tau=0.5$ sit near the
middle of their respective sweeps and perform comparably to every other tested
value at every density: the gate and the inverse-variance rule are
complementary, overlapping safeguards, and overall accuracy is robust to both
hyperparameters across more than an order of magnitude.

\begin{figure}[t]
\centering
\includegraphics[width=\linewidth]{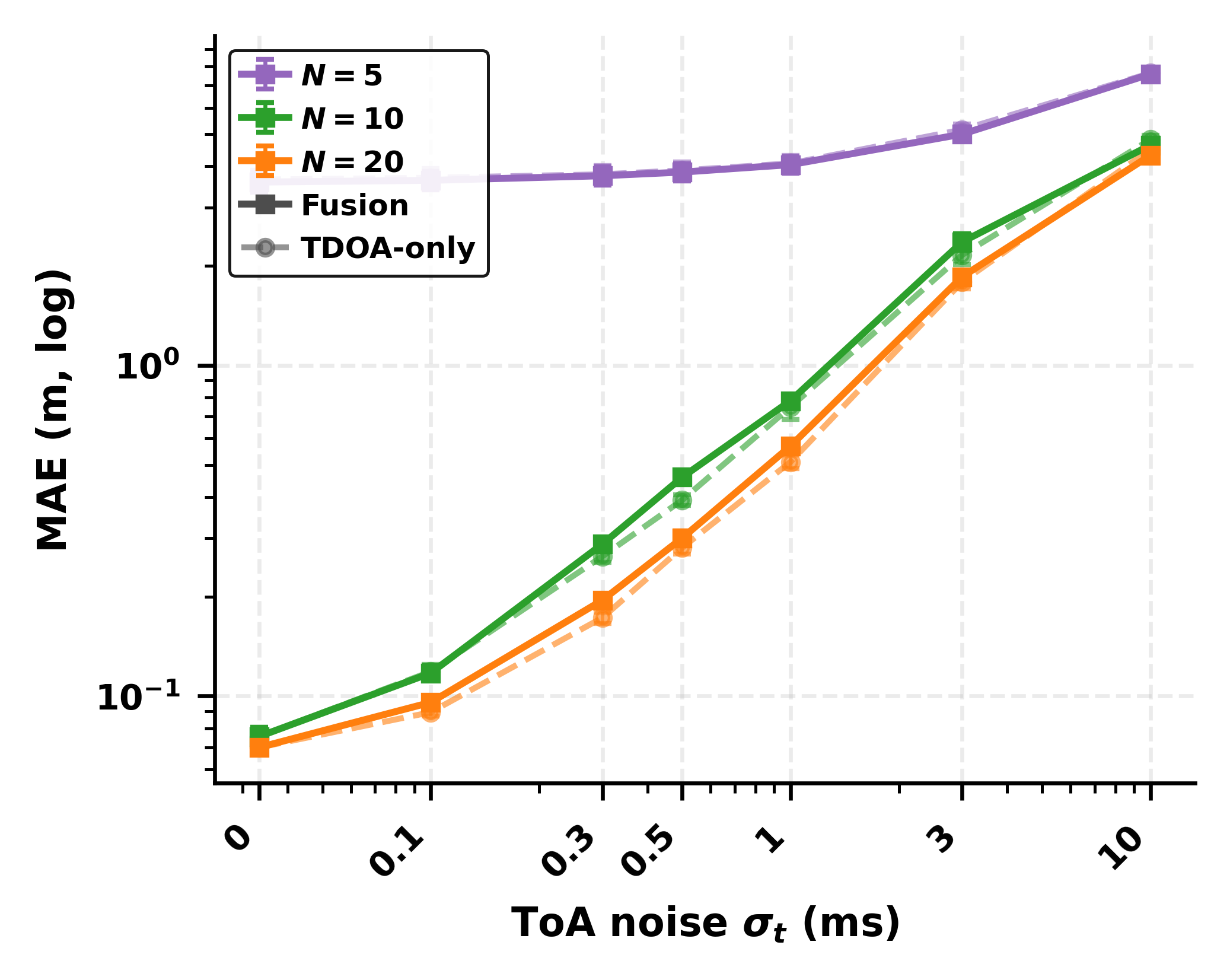}
\caption{TDOA-only and fusion MAE (m, mean $\pm$ 95\% CI, log scale) as a
function of ToA noise standard deviation $\sigma_t$ at $N \in \{5, 10, 20\}$.}
\label{fig:toanoise}
\end{figure}

Timing error tells a different story (Fig.~\ref{fig:toanoise}). Both
TDOA-only and fused MAE increase monotonically with $\sigma_t$ at all three
densities, but stay close to the noiseless solution for $\sigma_t \leq
1$~\si{\milli\second}: at $N=10$, MAE grows from $0.08$~\si{\meter} at
$\sigma_t=0$ to $0.75$~\si{\meter} at $\sigma_t=1$~\si{\milli\second}, then
sharply to $4.81$~\si{\meter} at $\sigma_t=10$~\si{\milli\second}---two orders
of magnitude beyond the default and well past practical microcontroller ToA
precision. A fixed per-agent bias $\sigma_b$ degrades performance considerably
more steeply than this zero-mean jitter (Table~\ref{tab:bias}): at $N=10$,
fused MAE rises from $0.46$ to $0.65$ to $1.48$~\si{\meter} as $\sigma_b$ goes
from $0$ to $1$ to $3$~\si{\milli\second}, and at $N=20$ from $0.30$ to $0.44$
to $1.11$~\si{\meter}, because a fixed offset shifts the linearized TDOA
system (Eq.~\eqref{eq:tdoa_lin}) by a constant that does not average out
across trials. In both cases the fused estimate tracks TDOA-only closely
(within $0.05$~\si{\meter} for jitter, $0.03$--$0.08$~\si{\meter} for bias),
mildly worse at $N \in \{10,20\}$ for intermediate error levels---increased
TDOA scatter reduces $\sigma^2_{\text{tdoa}}$'s advantage in
Eq.~\eqref{eq:fusion_weight}, raising $w_{\text{voc}}$ enough to occasionally
pull the fused estimate away from an already-reasonable TDOA solution---but
recovers a small, consistent advantage over TDOA-only at the most extreme
noise level ($\sigma_t=10$~\si{\milli\second}: $4.63$ vs.\
$4.81$~\si{\meter} at $N=10$), as VOC regains relative value once the acoustic
anchor itself becomes unreliable.

\begin{table}[t]
\centering
\caption{Fusion MAE (m, mean $\pm$ 95\% CI) across the per-agent clock bias
$\sigma_b$ sweep (on top of the default $\sigma_t = 0.5$~\si{\milli\second}).
The default $\sigma_b = 0$ is underlined.}
\label{tab:bias}
\footnotesize
\begin{tabular}{c ccc}
\toprule
$\sigma_b$ (ms) & $N{=}5$ & $N{=}10$ & $N{=}20$ \\
\midrule
$\underline{0}$   & $\underline{3.83{\pm}0.24}$ & $\underline{0.46{\pm}0.02}$ & $\underline{0.30{\pm}0.02}$ \\
$0.5$ & $3.85{\pm}0.23$ & $0.51{\pm}0.02$ & $0.35{\pm}0.02$ \\
$1$   & $3.93{\pm}0.23$ & $0.65{\pm}0.03$ & $0.44{\pm}0.02$ \\
$2$   & $4.20{\pm}0.23$ & $1.01{\pm}0.06$ & $0.71{\pm}0.03$ \\
$3$   & $4.42{\pm}0.22$ & $1.48{\pm}0.09$ & $1.11{\pm}0.05$ \\
\bottomrule
\end{tabular}
\end{table}

The default $\sigma_t=0.5$~\si{\milli\second} sits comfortably in the
low-sensitivity regime at all densities, confirming the headline results are
not an artifact of favorable noise. This corresponds to a range error of
$c\,\sigma_t \approx 17$~\si{\centi\meter}, well within the regime above.
However, because uncalibrated per-node clock offsets are markedly
more consequential than zero-mean jitter of the same magnitude, accurate
synchronization---which corrects fixed offsets rather than merely
bounds their variance---remains an important practical requirement.
\section{Conclusion}
\label{sec:conclusion}
This paper presented the first systematic study of acoustic, VOC, and
multimodal stress source localization within the Internet of Plants (IoP)
framework. TDOA multilateration gives an instantaneous, wind-independent,
sub-meter-accurate estimate once three or more agents are within acoustic
range, reaching a mean MAE of $0.46$~\si{\meter} ($\mathrm{SR}_{0.75}=0.81$) at
$N=10$ and $0.15$~\si{\meter} ($\mathrm{SR}_{0.75}=0.99$) at $N=50$ across the
52 evaluated scenarios. VOC-only localization is markedly less accurate, owing
to the narrow, advection-dominated geometry of the plume. Multimodal
fusion---an inverse-variance combination of the two estimates gated by VOC
informativeness---tracks the TDOA-only estimate closely, improving on both
single-modality baselines in 29 of 52 scenarios at $N=10$, with the corner
source the only persistent exception. Robustness analyses further showed this
performance floor to be stable under physical parameter uncertainty, ToA
noise and bias, the VOC gate threshold, and the Stage~1 bounding radius, with
fixed per-agent clock offsets identified as the dominant remaining source of
degradation. TDOA localization is deployable with currently available
acoustic hardware up to a moderate detection floor, while VOC localization
remains forward-looking pending compact, sub-ppb chemical sensors. To support
this and related work, the physics-based simulation dataset, the
finite-volume VOC solver, and the ray-based acoustic model are released as
open-source tools.

To move toward this goal, the framework established here can be extended in
several directions. The $15 \times 20$~\si{\meter} domain and constant,
uniform wind field studied here represent a single plot within a larger
open-field deployment, under laminar conditions; extending the simulation to
larger fields, multiple simultaneous sources, and temporally varying or
turbulent wind---which would widen the VOC plume and could improve its
localizability---are natural next steps, as is wind-aware agent placement.
Experimental validation in a physical planting environment is the most direct
path to confirming these simulation-based results, and would also allow the
ecological and agronomic impact of attached sensors on the designated agent
plants to be characterized empirically. On the hardware side, realizing VOC
localization in practice depends on continued progress in compact, low-power
chemical sensors reaching the sub-ppb sensitivities assumed here, while
tighter network-level time synchronization would further reduce the
clock-bias sensitivity identified above.

\bibliographystyle{IEEEtran}
\bibliography{references}
\vspace{-1cm}
\begin{IEEEbiography}
[{\includegraphics[width=1in, height=1.25in, clip, keepaspectratio]{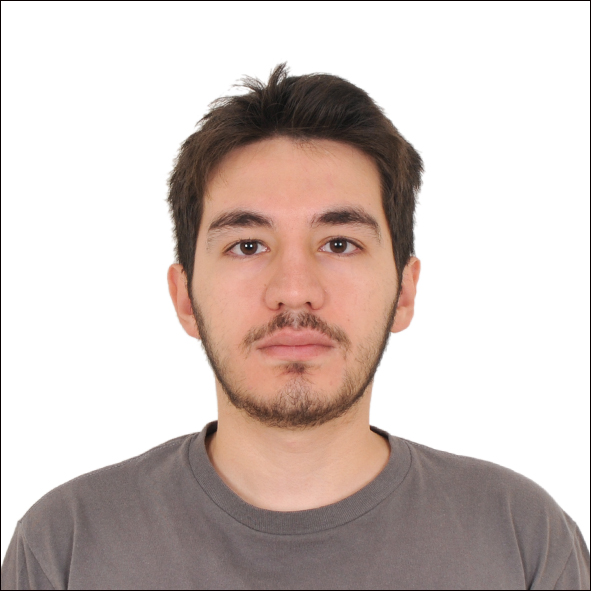}}]{Ahmet Burak Kilic} received the B.Sc. degree in Electrical and Electronics Engineering and the B.A. degree in Business Administration from Koç University, Istanbul, Turkey. He is currently pursuing the M.Sc. degree in Electrical and Electronics Engineering at Koç University under the supervision of Prof. Akan.
\end{IEEEbiography}
\vspace{-15cm}
\begin{IEEEbiography}
[{\includegraphics[width=1in,height=1.25in,clip,keepaspectratio]{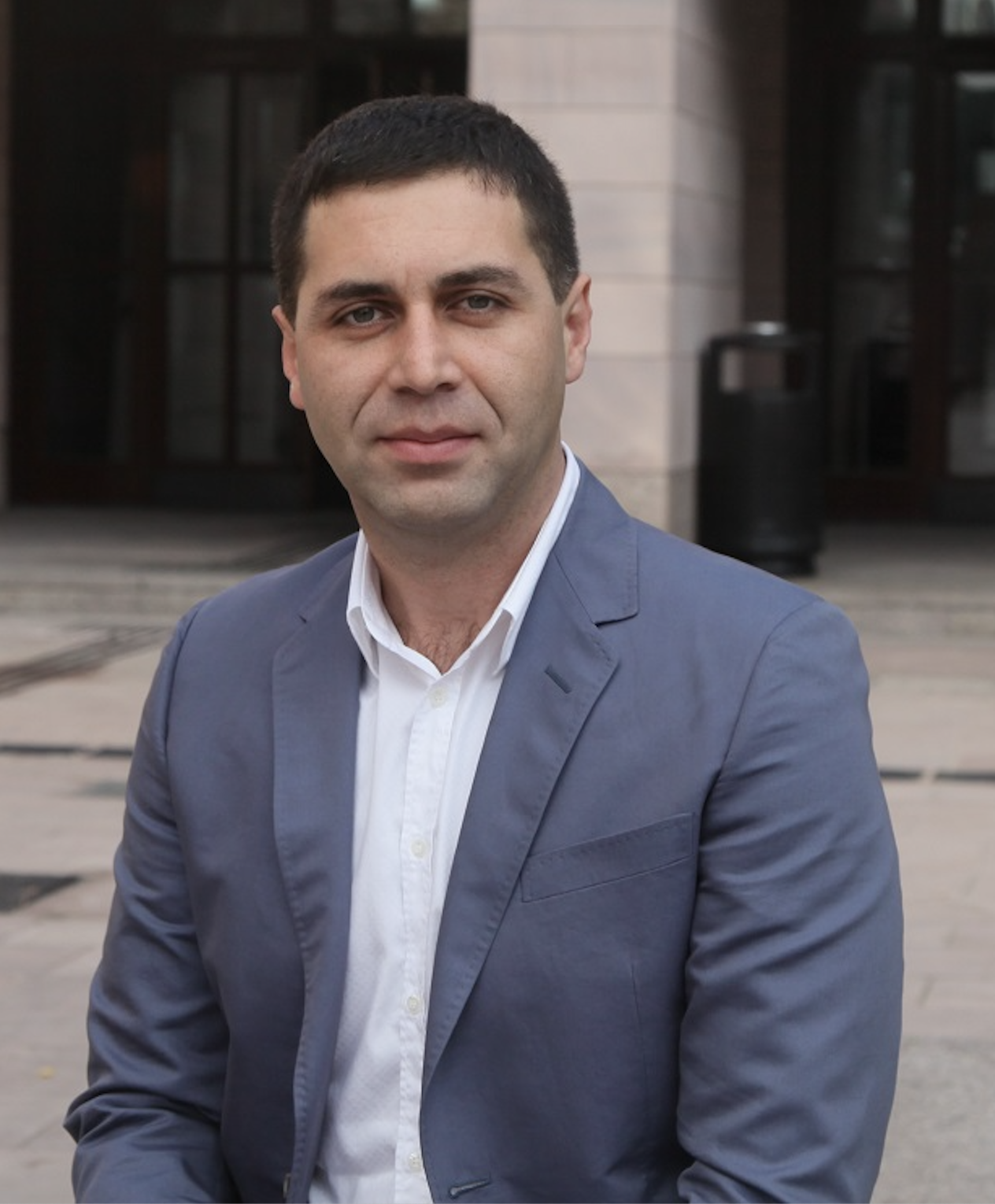}}]{Ozgur B. Akan (Fellow, IEEE)}
received the Ph.D. degree from the School of Electrical and Computer Engineering, Georgia Institute of Technology, Atlanta, in 2004. He is currently the Head of Centre for neXt Communications (CXC) at the Department of Engineering, University of Cambridge, UK and the Director of Centre for neXt-generation Communications (CXC), Koç University, Turkey. His research interests include wireless, nano, and molecular communications and Internet of Everything.
\end{IEEEbiography}
\end{document}